\documentclass[12pt]{article}
\usepackage{graphicx} 

\usepackage[small]{titlesec} 
\usepackage[utf8]{inputenc} 
\usepackage[T1]{fontenc}
\usepackage[english]{babel}
\usepackage[left=2cm,right=2cm,top=2cm,bottom=2cm,bindingoffset=0cm]{geometry}

\graphicspath{{figures/}}

\usepackage{setspace}
\usepackage{subcaption}
\usepackage{youngtab}
\usepackage{epigraph}
\usepackage{upgreek}

\usepackage[labelfont=bf, margin=20pt,font={small}]{caption}

\usepackage{amsmath}
\usepackage{esint}
\usepackage{bm}
\usepackage{
    tikz,
	amsmath,
	amssymb,
	amsfonts,
	amsthm,
	amscd,
	comment,
	epsfig,
	color,
	setspace,
	bbold,
	verbatim,
	slashed
}
\usepackage{upgreek}
\usepackage{tikz}
\usetikzlibrary{arrows}
\usepackage{mathrsfs}
\usepackage{multirow}
\usepackage{simpler-wick}
\usepackage{cite}

\newcommand\beq{\begin{equation}}
\newcommand\eeq{\end{equation}}

\tikzset{cross/.style={cross out, draw=black, minimum size=2*(#1-\pgflinewidth), inner sep=0pt, outer sep=0pt},
	cross/.default={5pt}}
\usepackage{mathtools}
\usepackage{graphicx}
\usepackage{authblk}
\usepackage{bbold}
\usepackage[makeroom]{cancel}
\usepackage[hidelinks]{hyperref}

	\definecolor{linkcolor}{rgb}{0,0,1} 
	\definecolor{urlcolor}{rgb}{0,0,1} 
 
\hypersetup{pdfstartview=FitH, linkcolor=linkcolor,urlcolor=urlcolor,citecolor=urlcolor, colorlinks=true}

\usepackage{url}
\usepackage{graphicx}
\usepackage{amssymb}

\usepackage[utf8]{inputenc}
\usepackage{amsmath}



\title{Once more about radiation from uniformly accelerating charge}
\author[1,2]{E.~T.~Akhmedov}
\author[1]{M.~N.~Milovanova}
\affil[1]{\itshape Institutskii per, 9, Moscow Institute of Physics and Technology, 141700, Dolgoprudny, Russia}
\affil[2]{\itshape Academician Kurchatov Square, 1, NRC ''Kurchatov Institute'', 123182, Moscow, Russia}

\begin{document}

{\let\nezpage\relax\maketitle}

\begin{abstract}
    We consider an electric charge uniformly accelerating along $x$ direction and moving with constant velocity along $y$ direction. We show that in the co-accelerating along $x$ direction Rinder's frame this charge creates non-zero Poynting vector, which, however, does not lead to a non-vanishing flux through an infinitely distant surface. Furthermore, we show that in the laboratory Minkowski frame such a charge creates a stress energy flux that does not vanish at infinity. We interpret these observations as that while the Rindler's frame corresponds to the static zone around the charge, the Minkowski frame does contain the wave zone. We give detailed calculations and explanations concluding that uniformly accelerating charge does radiate.    
\end{abstract}


\section{Introduction}

The issue of radiation from a uniformly accelerating charge has been extensively discussed in the literature (see, e.g., \cite{Pauli}, \cite{Fulton}, \cite{Peierls}). However, it remains noteworthy that there is no universally accepted consensus on this matter among experts.

The problem is complex due to its high symmetry and unusual nature: eternally accelerating charges do not occur in nature, and the answer to the question may, in principle, depend on the initial conditions and the precise definition of classical radiation. To place the issue in a broader context, it is important to note that it is closely related to the following series of problems\footnote{At least some of these questions are interconnected through coordinate transformations between Rindler and Minkowski reference frames.}:

\begin{enumerate}
    \item Does a charge fixed in a gravitational field produce radiation? (According to the equivalence principle, such a charge is equivalent to a uniformly accelerating charge in Minkowski spacetime.) If radiation is produced, would it be observed by a remote observer fixed in the gravitational field? Furthermore, if radiation is produced, would it be detected by a remote observer who is free-falling within the gravitational field?

    \item Does a charge in free fall within a homogeneous gravitational field emit radiation? (According to the equivalence principle, such a charge is equivalent to a charge in inertial motion within Minkowski spacetime.) If radiation is emitted, would it be observed by a co-falling remote observer? Furthermore, if radiation is emitted, would it be detected by a remote observer fixed in the gravitational field? Now consider a charge in inertial motion within Minkowski spacetime: does such a charge emit radiation in an accelerating reference frame?

    \item Does a charge in free fall within an expanding Universe emit radiation as observed from another free-falling reference frame within the same Universe\footnote{The answer to this question for point-like sources of radiation is provided in \cite{Akhmedov:2010ah}.}?
    
    \item What are the answers to the above questions if, instead of point-like sources of radiation, one considers, for example, scalar waves as the sources\footnote{Question No. 3 regarding the use of waves as sources of radiation for scalar fields was addressed in \cite{Akhmedov:2009be}.}? If radiation is produced by a wave as a source, while no radiation arises from a point-like source, what is the explanation for this apparent paradox?
\end{enumerate}
While some of these questions may have straightforward answers, others require a more thorough examination. In this paper, we present our analysis of the situation.

We will now outline several key facts, some of which are straightforward, while others are more complex:
\begin{itemize}
    \item Radiation is not merely a non-zero flux of the Poynting vector; it is a flux that does not vanish at infinity.
    \item The radiation friction force can only be defined after an averaging procedure and cannot be defined for a trajectory that lacks finite spatial support \cite{Peierls}. Therefore, the absence of radiation friction force does not necessarily imply the absence of radiation.
    \item The intensity of radiation, defined as the integral over a distant surface of the scalar product of the Poynting vector and the unit area vector, is not Lorentz invariant. The Lorentz invariant quantity is the power associated with the energy loss of the radiation source. This power, being independent of the reference frame, does not depend on the observer. If the power of energy loss is nonzero, radiation must be present.
    \item For a uniformly accelerating charge, the energy loss power is proportional to the square of the four-acceleration and, therefore, is nonzero.
    \item In the paper \cite{Kalinov}, it is demonstrated that the intensity of radiation from a uniformly accelerating charge in Minkowski space is nonzero at infinity. Furthermore, \cite{Kalinov} shows that, in Rindler coordinates, both the magnetic field and, consequently, the Poynting vector vanish.
    \item The electromagnetic field surrounding an accelerating charge forms two distinct regions: a close, static zone and a distant radiation zone. The static zone exists near the source, at distances shorter than the so-called radiation creation length, which is the characteristic distance the source must travel to produce radiation. In contrast, the radiation zone extends over distances much greater than the radiation creation length.
    \item For a uniformly accelerating charge with acceleration $a$, the radiation creation length (and the characteristic wavelength of the radiation) is $1/a$, assuming the speed of light is set to unity, $c=1$. The Rindler coordinates cannot be extended beyond a distance of $1/a$ from the charge in the direction of radiation.
\end{itemize}

The situation considered in \cite{Kalinov} is quite restrictive. To explore a more general case, this paper examines a charge that not only undergoes uniform acceleration along the first direction but also moves with constant velocity along the second spatial direction. This approach allows us to observe that, in a less restrictive scenario, the magnetic field in the co-accelerating frame is nonzero. Consequently, we find that the Poynting vector is nonzero in the co-moving frame. However, this Poynting vector does not correspond to any radiation. These observations lead us to classify the co-moving reference frame as being within the static zone. In contrast, radiation is present beyond the Rindler frame, in the wave zone.

\section{Set up of the problem}

We look for solutions of the Maxwell equations

\begin{align} \label{eom}
    \frac{1}{\sqrt{-g}} \partial_{\mu} (\sqrt{-g} F^{\mu \nu}) = 4 \pi j^{\nu}.
\end{align}
in two different reference frames of flat space-time --- the Minkowski metric:
\begin{align}
    ds^2 = dt^2 - dx^2 - dy^2 - dz^2
\end{align}
and the Rindler one:
\begin{align}
    ds^2 = \rho^2 d\tau^2 - d\rho^2 - dy^2 - dz^2.
\end{align}
The relation between these coordinate systems is as follows:
\begin{align} \label{coord_trans}
    t = \rho \sinh (\tau), \ x = \rho \cosh (\tau), \ y = y, \ z = z.
\end{align}
We use the Lorenz gauge:
\begin{align} \label{gauge}
    \partial^{\mu} A_{\mu} = g^{\mu \nu} \Gamma^{\eta}_{\mu \nu} A_{\eta}.
\end{align}
In the Maxwell equation we consider the source, $j_\mu$, created by a particle that is uniformly accelerating. 
Namely, we assume that the source is moving with constant 4-acceleration $a_{\mu} a^{\mu} = - a^2$ along the world-line parameterized as follows:
\begin{align} \label{worldline}
    z^{\alpha} (\theta) = \left( \frac{1}{a} \sinh (a\theta) \cosh (\phi), \ \frac{1}{a} \cosh (a\theta) , \ \frac{1}{a} \sinh (a\theta) \sinh (\phi), \ 0 \right), \quad a_\mu = \frac{d^2 z_\mu}{d\theta^2}
\end{align}
where $\theta$ --- its proper time. Such a world-line is obtained from the standard one 

\begin{align}
    \bar{z}^{\alpha} (\theta) = \left( \frac{1}{a} \sinh (a\theta), \ \frac{1}{a} \cosh (a\theta) , \ 0, \ 0 \right), 
\end{align}
by boosting the later along the $y$ axis with the rapidity $\phi$.
That is, we consider the source that is accelerating along the $x$ axis and moving with constant velocity along the $y$ one.

\section{The solution in Rindler's frame}

The easiest way to find the electromagnetic four-potential is to use the expressions for the Liénard-Wiechert potentials in Minkowski space-time for a charge moving along a given world-line \cite{Lienard, Wiechert}:
\begin{align} \label{Liénard}
      A_\mu (t, \vec{x}) = \frac{\left(1, -\vec{v}\right)}{\left| \vec{R} \right| - \left( \vec{v}, \vec{R} \right) }, \quad {\rm where} \quad
      \big| \vec{R} \big| = t - t_{rad},
\end{align}
in our case $t_{rad} = z_0(\theta, \phi)$ and $\vec{R} = \vec{x} - \vec{z}(\theta, \phi)$. 

The relation between $R$ and $t$ establishes that the proper time of the moment of radiation and the time of the detection of the radiation are connected by the light-like world-line 
and the three-speed of the source $\vec{v} (\theta,\phi) = d \vec{z} (\theta,\phi)/d z^{0} (\theta,\phi)$ is equal to:
\begin{align}
    \vec{v} (\theta, \phi) = \left( \frac{\tanh (a \theta)}{\cosh (\phi)}, \ \tanh (\phi), \ 0 \right).
\end{align}

Squaring the relation between $R$ and $t$ and substituting into it the world-line of the source (\ref{worldline}), we find the relation between the event of the detection of the radiation and the event of its radiation
\begin{align}\label{txyz1a}
    & -t^2 + x^2 + y^2 + z^2 + \frac{1}{a^2} - \frac{2 y}{a} \sinh (a\theta) \sinh (\phi) = \nonumber \\
    & = \frac{2 x}{a} \cosh (a\theta) - \frac{2 t}{a} \sinh (a\theta) \cosh (\phi),
\end{align}
and in the Rindler region
\begin{align}\label{txyz1a_rindler}
    & \frac{1}{a^2} + \rho^2 + y^2 + z^2 = \nonumber \\
    & = \frac{2}{a} \Big(\rho \cosh (\tau) \cosh (a\theta)  + y \sinh (a\theta) \sinh (\phi) - \rho \sinh (\tau) \sinh (a\theta) \cosh (\phi)\Big),
\end{align}
which we will use below.

Then let us write the denominator of the right hand side of eq.(\ref{Liénard}) in Minkowski coordinates:
\begin{align}
\begin{split}
    \big| \vec{R} \big| - \left( \vec{v}, \vec{R} \right)
    = t - \frac{\tanh (a \theta)}{\cosh (\phi)} x - \tanh (\phi) y,
\end{split}
\end{align}
and in Rindler ones
\begin{align}
\begin{split}
    \big| \vec{R} \big| - \left( \vec{v}, \vec{R} \right) = \rho \sinh (\tau) - \frac{\rho \cosh (\tau) \tanh (a \theta)}{\cosh (\phi)} - \tanh (\phi) y,
\end{split}
\end{align}
where we have used the relation between $t_{rad}$ and $\big| \vec{R} \big|$ from (\ref{Liénard}). These relations will be also used below.

We want to find and compare the expressions for the Lienard-Wiechert potentials and electromagnetic fields in the Rindler and Minkowski coordinate frames. They are related by the coordinate transformation from (\ref{coord_trans}) as components of the four-vector potential and the electromagnetic tensor, respectively.

Before writing out the expression for the four-potential in the Rindler's frame, let us use the remaining gauge freedom to simplify it. The Lorenz gauge fixing condition \eqref{gauge} in this frame is as follows:
\begin{align}
    g^{\mu \nu} \partial_{\mu} A^R_{\nu} = \frac{A^R_1}{\rho},
\end{align}
where $g^{\mu\nu}$ is the inverse Rindler metric and we have introduced the index $R$ just to stress that these expressions are attributed to the Rindler's frame, to distinguish them from those in the Minkowski frame (without an index).
Then the scalar function $\alpha(\tau, \rho, y, z)$ performing the remaining gauge freedom, i.e. the gauge transformation $A_\mu - \partial_\mu \alpha$ that does not change the gauge condition, must solve
\begin{align}
    g^{\mu \nu} \partial_{\mu} \partial_{\nu} \alpha = \frac{\partial_{1} \alpha}{\rho}.
\end{align}
This equation has such a solution as $\alpha = C \cdot \ln (\rho)$ with an arbitrary constant $C$. Let us use such a remaining gauge transformation over the first component of the four-potential, $A^{R}_{1} = A^{R}_{1} - \rho^{-1}$, to change its expression. Then the electromagnetic four-potential in Rindler coordinates takes the following form:
\begin{align} \label{a_in_rindler}
      A_{0}^{R} = \frac{\rho  \Big[\cosh (\tau -a \theta ) + \cosh (\tau ) \cosh (a \theta ) \Big(\cosh (\phi )-1\Big)\Big]}{f(\tau, \rho, y, z, \theta)}, \nonumber \\
      A_{1}^{R} = \frac{y}{\rho} \cdot \frac{ \cosh (a \theta ) \sinh (\phi )}{f(\tau, \rho, y, z, \theta)}, \nonumber \\
      A_{2}^{R} = -\frac{\cosh (a \theta ) \sinh (\phi ) }{f(\tau, \rho, y, z, \theta)}, \nonumber \\
      A_{3}^{R} = 0.
\end{align}
To simplify these expressions, we introduce the following function:
\begin{align} \label{function}
    & f(\tau, \rho, y, z, \theta) = \rho \sinh (\tau -a \theta ) - \cosh (a \theta ) \Big[y \sinh (\phi ) + \rho \sinh (\tau ) \Big(1 - \cosh (\phi )\Big)\Big].
\end{align}
Note that in the Rindler frame $A^R_0$ has a different dimensionality from $\vec{A}^R$. That is because while $\rho, \, y, \,z$ have dimension of length, the time coordinate $\tau$ is dimensionless. For the same reason, the electric field, $\vec{E}^R$, also has a different dimensionality from the magnetic one, $\vec{B}^R$.

The corresponding electric and magnetic fields in the Rindler frame are as follows:
\begin{align}
      E_{1}^{R} & = \frac{a \rho \Big[ 2 y \rho \sinh (\tau ) \sinh (\phi )- \cosh (\phi ) \Big( -\rho ^2+y^2 +z^2+a^{-2} \Big) \Big] }{2 f(\tau, \rho, y, z, \theta)^3}, \nonumber \\
      E_{2}^{R} & = -\frac{ a \rho \left[ \sinh (\tau ) \sinh (\phi ) \Big(\rho ^2 - y^2 + z^2 + a^{-2}\Big)-2 \rho  y \cosh (\phi ) \right]}{2 f(\tau, \rho, y, z, \theta)^3} , \nonumber \\
      E_{3}^{R} & = \frac{ a \rho z \Big[\rho  \cosh (\phi )+y \sinh (\tau ) \sinh (\phi )\Big]}{f(\tau, \rho, y, z, \theta)^3}.
\end{align}
and
\begin{align}
      & B_{1}^{R} = \frac{a \rho z \cosh (\tau ) \sinh (\phi )}{f(\tau, \rho, y, z, \theta)^3}, \nonumber \\
      & B_{2}^{R} = \frac{a y z \cosh (\tau ) \sinh (\phi )}{f(\tau, \rho, y, z, \theta)^3}, \nonumber \\
      & B_{3}^{R} = \frac{ a \left( -\rho ^2-y^2+z^2 +a^{-2}\right) \cosh (\tau ) \sinh (\phi )}{2 f(\tau, \rho, y, z, \theta)^3}.
\end{align}
Then the corresponding Poynting vector is:
\begin{align}\label{PoyntingRind}
      & S_{1}^{R} = \frac{a^2 \rho  \cosh (\tau ) \sinh (\phi )}{16 \pi f(\tau, \rho, y, z)^6} \, \Bigg[-2 \rho  y \cosh (\phi ) \Big( \rho ^2+z^2+y^2 - a^{-2}\Big) - \nonumber \\
      & - \sinh (\tau ) \sinh (\phi ) \Big( -\rho ^4+y^4+z^4+2 y^2 z^2 +2 a^{-2} \left(z^2-y^2\right) + a^{-4}\Big) \Bigg], \nonumber \\
      & S_{2}^{R} = \frac{a^2 \rho  \cosh (\tau ) \sinh (\phi )}{16 \pi f(\tau, \rho, y, z)^6} \, \Big[2 \rho  y \sinh (\tau ) \sinh (\phi ) \Big( \rho^2+y^2+z^2 - a^{-2}\Big) + \nonumber \\
      & + \cosh (\phi ) \Big( \rho ^4-y^4+z^4+2 \rho ^2 z^2 +2 a^{-2} \left(z^2-\rho ^2\right) +a^{-4} \Big) \Big], \nonumber \\      
      & S_{3}^{R} = \frac{a^2 \rho z \cosh (\tau ) \sinh (\phi ) }{8 \pi f(\tau, \rho, y, z)^6} \, \Big( \rho ^2+y^2+z^2 +a^{-2}\Big) \Big[\rho  \sinh (\tau ) \sinh (\phi )-y \cosh (\phi )\Big] .
\end{align}
When we return to the situation that was considered in \cite{Kalinov}, $\phi=0$, we reproduce the results of that paper. In the case $\phi=0$ the magnetic field and Poynting vector in the Rindler's frame are zero. Meanwhile in our situation these quantities are not zero and the question of the presence or the absence of the radiation in the Rindler's frame demands a careful consideration.

To answer the last question we have to find the behavior of the intensity of the radiation (stress-energy flux) within Rindler's region but far away from the source. Using the relation between $t_{rad}$ and $\big| \vec{R} \big|$ from (\ref{Liénard}) the distance from the source $\big|\vec{R}\big|^2$, 
can be written in Rindler's coordinates as:
$$ 
\left|\vec{R}\right|^2 = \Big( \rho \sinh (\tau) - a^{-1} \sinh(a \theta) \cosh(\phi) \Big)^2.
$$
In has to be taken to infinity in units of acceleration. Essentially our goal is to single out in (\ref{PoyntingRind}) the last function of $\tau, \rho, y, z$ in the limit $\big|\vec{R}\big|^2 \to \infty$. But there are different ways to go far away from the source of radiation within the Rindler's wedge. Let us examine them separately. 

To fulfill the relation \eqref{txyz1a_rindler} in the limit $\big|\vec{R}\big|^2 \to \infty$, the Rindler's time $\tau$ should also be taken to infinity, which can be expected on general grounds. Let us take the large distance limit in such a way that $\rho \sim e^{k\tau}, k \in \left[0, 1 \right]$. Note that $k$ cannot be greater than 1, since then the relation \eqref{txyz1a_rindler} will not be satisfied. Then there are several ways in which the spatial coordinates tend to infinity:

1. If $k=1$, then $y, z$ can be taken to be of the order of $\rho$ in the limit in question, or they can be taken to remain finite, i.e. will not tend to infinity;

2. If $k<1$, then we should assume that $y^2 + z^2 \sim e^{(k+1)\tau}$ to fulfill \eqref{txyz1a_rindler}.

In any case the term $y \sinh(a \theta) \sinh(\phi)$ on the right hand side of \eqref{txyz1a_rindler} is much smaller than all the other terms. Therefore, we can drop this term off.
In all, we consider such a limit, in which: 

\begin{align} \label{BigDistanceRelation}
    \cosh (\tau) \approx \sinh (\tau) \approx \frac{e^\tau}{2} \approx \frac{\rho^2 + y^2 + z^2}{2 a^{-1} \rho \left( \cosh (a \theta) - \sinh(a \theta) \cosh(\phi) \right)}.
\end{align}
Substituting these expressions into \eqref{PoyntingRind} gives the approximate expression for the components of the Poynting vector at large spatial distances within the Rindler's wedge:
\begin{align}
      & S_{1}^{R} \approx \frac{a^{-2} \sinh^2 (\phi ) \Big( \cosh (a \theta) - \sinh(a \theta) \cosh(\phi) \Big)^4}{ \pi \rho \Big( \rho ^2+y^2+z^2\Big)^4 \Big(\sinh(a \theta) - \cosh(a \theta) \cosh (\phi ) \Big)^6} \times \nonumber \\
      & \times \Big[- 4  a^{-1} \rho^2 y \coth (\phi ) \Big( \cosh (a \theta) - \sinh(a \theta) \cosh(\phi) \Big) + \rho ^4 - \left(y^2+z^2\right)^2 \Big], \nonumber \\
      & S_{2}^{R} \approx \frac{2 a^{-2} y \sinh^2 (\phi ) \Big( \cosh (a \theta) - \sinh(a \theta) \cosh(\phi) \Big)^4}{\pi \Big( \rho ^2+y^2+z^2\Big)^3 \Big(\sinh(a \theta) - \cosh(a \theta) \cosh (\phi ) \Big)^6}, \nonumber \\
      & S_{3}^{R} \approx \frac{2 a^{-2} z \sinh^2 (\phi ) \Big( \cosh (a \theta) - \sinh(a \theta) \cosh(\phi) \Big)^4}{\pi \Big( \rho ^2+y^2+z^2\Big)^3 \Big(\sinh(a \theta) - \cosh(a \theta) \cosh (\phi ) \Big)^6},
\end{align}
as follows from \eqref{BigDistanceRelation} in the limit under consideration $\rho^2 + y^2 + z^2 \rightarrow \infty$ in units of the acceleration $1/a$. This then leads to the following expression for the energy flux through a solid angle $d \Omega$ at long distance from the source (within the Rindler's wedge):
\begin{align}
    & \frac{dI^{R}}{d \Omega} = \Big(\vec{S}^R, \vec{n} \Big) \left|\vec{R}\right|^2 \sim \frac{\sinh^2\left(\phi\right)}{2 \rho \Big( \rho ^2+y^2+z^2 \Big)^2} \, \times \nonumber \\
    & \times \bigg(\rho^4 - (y^2 + z^2)^2 - 4 \rho a^{-1} (\rho y - (y^2 + z^2)) (\cosh (a \theta) - \sinh(a \theta) \cosh(\phi) ) \bigg).
\end{align}
This expression is vanishing for the both ways to take the large spatial distance limit, which have been described above, except $k = 0$. In the latter case the flux in the infinity is not zero in some directions and even sometimes is negative. We give technical details in the appendix \ref{app_A}. Please note that the resulting flux is zero in the limit $\phi = 0$. And what is even more important is that the flux through the infinity depends on the spatial coordinate $\rho$, which is never the case in standard situations. Perhaps that is the hint for the resolution of the puzzle: the notion of the large spatial distance limit in the Rindler's frame is not that straightforward. 

\section{The solution in Minkowski frame}

Let us now consider the situation in the Minkowski frame. We can either directly calculate the electromagnetic fields from (\ref{Liénard}) or transform back to the Minkowski frame the fields found above in Rindler's frame.
As an independent check both ways we obtain the same answer.

Using eq. (\ref{txyz1a}) and eq. (\ref{coord_trans})
to express $t$ and $x$ via Rindler's coordinates, we obtain    
\begin{align} \label{connection}
    \cosh (\tau - a \theta) = \nonumber \\
    = \frac{a}{2 \rho } \, \Big[ \frac{1}{a^2}+\rho ^2+y^2+z^2 - \frac{2 \, \sinh (a \theta )}{a} \Big( y \sinh (\phi ) + \rho \sinh (\tau ) \Big[1-\cosh (\phi )\Big]\Big) \Big] 
\end{align}
In the following expressions this relationship will be implied wherever there the functions $\cosh (\tau - a \theta)$ and $\sinh (\tau - a \theta)$ are present.
Then, making the change of the coordinates to the Minkowskian coordinates in \eqref{function} we obtain the functions:
\begin{align}
\begin{split}
    & f(t, \vec{x}) = - \cosh (a \theta ) \Big( y \sinh (\phi ) + t (1 - \cosh (\phi ) \Big) + \\
    & +\frac{1}{2} \sqrt{4 \Big(t^2-x^2\Big) + \Big( a \Big(a^{-2}-t^2+x^2+y^2+z^2\Big)-2 \sinh (a \theta ) \Big[ y \sinh (\phi ) + t \Big(1-\cosh (\phi ) \Big) \Big] \Big)^2 },
\end{split}
\end{align}
where we have used (\ref{connection}) to express out $\sinh (\tau - a \theta)$.

Then, the electric and magnetic fields in the Minkowski frame can be written as follows:
\begin{align} \label{ElMink}
      & E_{1} = -\frac{a}{2 f(t, \vec{x})^3} \Big( \cosh (\phi ) \Big(a^{-2} + t^2 - x^2+y^2+z^2 \Big) - 2 y t \sinh (\phi ) \Big), \nonumber \\
      & E_{2} = \frac{a x \Big(y \cosh (\phi )-t \sinh (\phi ) \Big)}{f(t, \vec{x})^3}, \nonumber \\
      & E_{3} = \frac{a x z \cosh (\phi )}{f(t, \vec{x})^3}.
\end{align}
and
\begin{align} \label{BMink}
      & B_{1} = \frac{a x z \sinh (\phi )}{f(t, \vec{x})^3}, \nonumber \\
      & B_{2} = \frac{a z \Big(y \sinh (\phi )-t \cosh (\phi ) \Big)}{f(t, \vec{x})^3}, \nonumber \\
      & B_{3} = \frac{a}{2 f(t, \vec{x})^3} \Big( \sinh (\phi ) \Big(a^{-2}- t^2-x^2-y^2+z^2\Big) + 2 y t \cosh (\phi ) \Big).
\end{align}
And the resulting Poynting vector is:

\begin{align} \label{PoyntMink}
      & S_{1} = \frac{a^2 x}{8 \pi f(t, \vec{x})^6} \Big(2 t y \sinh (\phi ) \Big( y \sinh (\phi ) - t \cosh (\phi ) \Big) +2 t \left(y^2+z^2\right) - \nonumber \\
      &  + \sinh (\phi ) \Big( t \sinh (\phi ) - y \cosh (\phi ) \Big) \Big( t^2+x^2+y^2+z^2 -a^{-2} \Big) \Big), \nonumber \\
      & S_{2} = \frac{a^2}{32 \pi f(t, \vec{x})^6} \Big( 8 t y \cosh^2 ( \phi )  \left(t^2+y^2\right) - 4 t y \left(t^2+x^2+y^2-z^2 -a^{-2}\right)+ \nonumber \\
      & + 2 \sinh ( \phi ) \cosh ( \phi ) \Big( \left(z^2 - x^2 + a^{-2} \right)^2 - \left( t^2+y^2 \right)^2 +4 \left(x^2 z^2 - t^2 y^2 \right) \Big) \Big), \nonumber \\
      & S_{3} = \frac{a^2 z}{8 \pi f(t, \vec{x})^6} \Big( - y \sinh (\phi ) \cosh (\phi ) \Big( 3 t^2+x^2+y^2+z^2 +a^{-2}\Big) + \nonumber \\
      & +t \cosh ^2(\phi ) \Big( t^2+x^2+3 y^2+z^2 +a^{-2}\Big)- 2 t \left(x^2+y^2\right) \Big) .
\end{align}
These expressions reduce to those found in \cite{Kalinov} in the limit $\phi = 0$. Furthermore, in that paper it was shown that the flux through the infinitely distant surface is not zero, when the radiation achieves the surface after the appropriate time, i.e. there is an energy flux emitted by the homogeneously accelerating charge which decouples from the source and can be measured by a very distant inertial observer in flat space-time. Obviously the same is true for the case when $\phi \neq 0$. But still let us be a bit more explicit and give some details.

Namely, let us show that the energy flux, which follows from (\ref{PoyntMink}), through an infinitely distant surface is not zero. To perform the integration over the distant surface, we choose the following parametrization:
\begin{align} \label{parametrization}
\begin{split}
    t = r, \ x = r \cos(\alpha), \ y = r \sin(\alpha) \cos (\beta), \ z = r  \sin(\alpha) \sin (\beta),
\end{split}
\end{align}
and take the limit $a r \rightarrow \infty$. In this parametrization
eq. (\ref{PoyntMink}) reduces to:
\begin{align}
      & S_{1} = \frac{ a^2 \cos (\alpha )}{8 \pi r^2 f(r, \alpha, \beta)^6} \Big( \sinh ^2(\phi ) \Big(2 +2 \sin ^2(\alpha ) \cos ^2(\beta )-(a r)^{-2} \Big)+ \nonumber \\
      & + \Big((a r)^{-2} -4 \Big) \sin (\alpha ) \cos (\beta ) \sinh (\phi ) \cosh (\phi )+2 \sin ^2(\alpha ) \Big), \nonumber \\
      & S_{2} = \frac{ a^2 }{16 \pi r^2 f(r, \alpha, \beta)^6} \Big( 2 \sin (\alpha ) \cos (\beta ) \Big( (a r)^{-2} +2 \sinh ^2(\phi ) \Big(\sin ^2(\alpha ) \cos ^2(\beta )+1 \Big)+2 \sin ^2(\alpha ) \Big)+ \nonumber \\
      & + \sinh (\phi ) \cosh (\phi ) \Big( (a r)^{-4} -2 (a r)^{-2} \left(\cos ^2(\alpha )-\sin ^2(\alpha ) \sin ^2(\beta )\right)-8 \sin ^2(\alpha ) \cos ^2(\beta ) \Big) \Big), \nonumber \\
      & S_{3} = \frac{ a^2 \sin (\alpha ) \sin (\beta )}{8 \pi r^2 f(r, \alpha, \beta)^6} \Big( \sinh ^2(\phi ) \Big((a r)^{-2} +2 \sin ^2(\alpha ) \cos ^2(\beta )+2 \Big)- \nonumber \\
      & - \Big((a r)^{-2}+4 \Big) \sin (\alpha ) \cos (\beta ) \sinh (\phi ) \cosh (\phi )+\Big( (a r)^{-2} +2 \sin ^2(\alpha ) \Big) \Big),
\end{align}
where
\begin{align}
\begin{split}
    & f(r, \alpha, \beta) = - \cosh (a \theta ) \Big( \sin(\alpha) \cos (\beta) \sinh (\phi ) + (1 - \cosh (\phi )) \Big) + \\
    & + \sqrt{\sin^2 (\alpha) + \Big( (2 a r)^{-1} - \sinh (a \theta ) \Big[ \sin(\alpha) \cos (\beta) \sinh (\phi ) + (1-\cosh (\phi ))\Big] \Big)^2 }.
\end{split}
\end{align}
Then the energy flux through a solid angle $d \Omega$ is:
\begin{align}
\begin{split}
    & \frac{dI}{d \Omega} = \Big( \vec{S}, \vec{n} \Big) r^2 = \frac{a^2 }{16 \pi f(r, \alpha, \beta)^6} \Big( (a r)^{-4} \sin (\alpha ) \cos (\beta ) \sinh (\phi ) \cosh (\phi ) + \\
    & + 2 (a r)^{-2} \Big( 1 + \sinh ^2(\phi ) \sin ^2(\alpha ) \sin ^2(\beta )- \cosh ^2(\phi ) \cos ^2(\alpha ) \Big)+ \\
    & + 4 \Big( \sin ^2(\alpha ) + \sinh ^2(\phi ) \left(\sin ^2(\alpha ) \cos ^2(\beta )+1\right) - \sin (\alpha ) \cos (\beta ) \sinh (2 \phi ) \Big) \Big),
\end{split}
\end{align}
where $\vec{n} = \left( \cos(\alpha), \sin(\alpha) \cos(\beta), \sin(\alpha) \sin(\beta) \right)$. In the large distance limit, $a r \to \infty$, it reduces to: 
\begin{align} \label{fluxz}
\begin{split} 
    \frac{dI}{d \Omega} \approx \frac{a^2 \Big[ \sin ^2(\alpha ) + \sinh ^2(\phi ) \Big(\sin ^2(\alpha ) \cos ^2(\beta )+1\Big) - \sin (\alpha ) \cos (\beta ) \sinh (2 \phi ) \Big]}{4 \pi g(\alpha, \beta)^6},
\end{split}    
\end{align}
where
\begin{align}
\begin{split}
    & g(\alpha, \beta) = - \cosh (a \theta ) \Big[ \sin(\alpha) \cos (\beta) \sinh (\phi ) + (1 - \cosh (\phi )) \Big] + \\
    & + \sqrt{\sin^2 (\alpha) + \sinh^2 (a \theta ) \Big[ \sin(\alpha) \cos (\beta) \sinh (\phi ) + (1-\cosh (\phi ))\Big]^2}. 
\end{split}
\end{align}
When $\phi=0$ the expression for the flux that we find here reduces to the one found in \cite{Kalinov}.

Note that the obtained expression is not zero. Thus, there is a flux through a distant surface. The expression (\ref{fluxz}) is always non-negative and, therefore, the total integrated flux is non-zero.

\section{Peculiar properties of the solution and other options}

In this section we consider several peculiar properties of the solutions of the Maxwell's equations which have been found above.
Consider, e.g., the limit in which the acceleration, $a$, of the source of the radiation is taken to zero. Then the electromagnetic four-potential in the Rindler's frame becomes a pure gauge:

\begin{align} \label{rindler_a_0}
    \lim_{a \rightarrow 0} A^{R}_{\mu} (\rho \rightarrow a^{-1} e^{\pm \tau}, \tau, y, z) = \left(\coth (\tau), 0, 0, 0 \right)
\end{align}
or
\begin{align}
    \lim_{a \rightarrow 0} A^{R}_{\mu} (\tau \rightarrow \pm \infty, \rho, y, z) = \left(\pm 1, 0, 0, 0 \right).
\end{align}
Detailed calculations can be found in the appendix \ref{app_B}. Therefore, in the absence of acceleration, the electric and magnetic fields are zero meanwhile the source seems to be still present, just does not accelerate. 

Furthermore, in the appendix \ref{app_B} we also show that in the Minkowski frame when $a\to 0$ the electromagnetic four-potential is not a pure gauge only in one of the cases: 
\begin{align}
    & \lim_{a \rightarrow 0} A_{\mu} (x \rightarrow a^{-1} \cosh(a \theta), t, y, z) = \left( \cosh(\phi), 0,\frac{\sinh(\phi)}{ \theta - t \cosh(\phi) + y \sinh(\phi)}, 0 \right), \nonumber \\
    & \lim_{a \rightarrow 0} A_{\mu} (t \rightarrow \pm a^{-1}, x, y, z) = \left( 0, 0, 0, 0 \right).
\end{align}
The fact that the electric and magnetic fields are not zero only when $x \rightarrow a^{-1} \cosh(a \theta)$ will be explained below. For all other cases both magnetic and electric fields tend to zero when the source does not accelerate. Thus, we seem to have a paradox: for the world-line under consideration non-accelerating charge, $a\to 0$, although being present, does not create even an electric field. 

The resolution of this apparent puzzle is as follows. In the limit $a\to 0$, the world-line of the charge (\ref{worldline}) is taken infinitely far away along one of the spatial directions:
\begin{align}
    \lim_{a \rightarrow 0} z^{\alpha} (\theta) = \left( \theta \cosh \phi, \ \infty , \ \theta \sinh \phi, 0 \right).
\end{align}
Therefore, only in the limit when $x \rightarrow a^{-1} \cosh(a \theta)$ in the Minkowski frame the electromagnetic fields are non-zero.

To clarify the situation let us consider a bit different world-line for which the position of the charged source does not get shifted infinitely far away in the limit of zero acceleration:
\begin{align}\label{newworldline}
    z^{\alpha} (\theta) = \left( \frac{1}{a} \sinh a\theta \cosh \phi, \ \frac{1}{a} \Big[\cosh (a\theta) - 1\Big], \ \frac{1}{a} \sinh a\theta \sinh \phi, 0 \right).
\end{align}
In such a case obviously
\begin{align}
    \lim_{a \rightarrow 0} z^{\alpha} (\theta) = \Big( \theta \cosh \phi, \ 0 , \ \theta \sinh \phi, 0 \Big).
\end{align}
However, this world-line does not fully (for all values of its proper time) lie in the Rindler's region that we consider. In fact, the condition for the world-line to fit into the appropriate quadrant, $| z^{0} | \leq z^{1} $, is fulfilled only when  
\begin{align}
      \cosh (\phi) \leq \frac{\cosh (a\theta) -1}{| \sinh (a\theta) |} \leq 1.
\end{align}
Let us find what electric and magnetic fields does the charge create in Minkowski frame if it moves along such a world-line as (\ref{newworldline}). The corresponding relation between the time of radiation and the time of the detection of the radiation in Minkowski frame is as follows:
\begin{align}
\begin{split}
    & x \cosh (a \theta) - t \sinh (a \theta) = x + \frac{a}{2} \Big( -t^2 + x^2 + y^2 + z^2 \Big) + \frac{1}{a} \Big( 1- \cosh (a \theta ) \Big) + \\
    & + \sinh (a \theta ) \Big(t (\cosh (\phi )-1) - y \sinh (\phi ) \Big). 
\end{split}
\end{align}
Then, for this world-line we have that:

\begin{align}
\begin{split}
    & \left| \vec{R} \right| - \left( \vec{v}, \vec{R} \right) = t - \frac{\sinh a \theta}{\cosh a \theta \cosh \phi} \left(x + a^{-1} \right) - \frac{\sinh \phi}{\cosh \phi} y.
\end{split}
\end{align}
In such a case the four-potential looks like:
\begin{align}
    A_{\mu} (t, x, y, z) = \frac{1}{f(t, x, y, z)} \Big( \cosh a \theta \cosh \phi, - \sinh a \theta, - \cosh a \theta \sinh \phi , 0 \Big)
\end{align}
where
\begin{align}
    & f(t, x, y, z) = \Big(t \cosh (a \theta) - x \sinh (a \theta) \Big) - a^{-1} \sinh a \theta + \nonumber \\
    & + \cosh a \theta \Big( t (\cosh \phi - 1)  - y \sinh \phi \Big).
\end{align}
Therefore, the electric and magnetic fields are
\begin{align}
      E_{1} & = \frac{a}{2 f(t, x, y, z)^3} \Big( 2 t y \sinh (\phi )-  \cosh (\phi ) \left( t^2-x^2+y^2+z^2-2 a^{-1} x\right) \Big), \nonumber \\
      E_{2} & = \frac{ (a x+1) \Big(y \cosh (\phi )-t \sinh (\phi ) \Big)}{f(t, x, y, z)^3}, \nonumber \\
      E_{3} & = \frac{ z (a x+1) \cosh (\phi )}{f(t, x, y, z)^3},
\end{align}
and
\begin{align}
      B_{1} & = \frac{z (a x+1) \sinh (\phi )}{f(t, x, y, z)^3} , \nonumber \\
      B_{2} & = -\frac{a z \Big(t \cosh (\phi )-y \sinh (\phi ) \Big)}{f(t, x, y, z)^3} , \nonumber \\
      B_{3} & = \frac{a}{2 f(t, x, y, z)^3} \Big(2 t y \cosh (\phi ) - \sinh (\phi ) \left(t^2+x^2+y^2-z^2+2 a^{-1} x\right) \Big) ,
\end{align}
In the case when $\phi=0$ these expressions reduce to:
\begin{align} \label{small_angle_e2}
       E_{1} & = -\frac{4 a \Big( t^2-x^2+y^2+z^2 -2 a^{-1} x \Big) }{\left( \sqrt{\Big( 2 x + a ( -t^2 + x^2 + y^2 + z^2 ) + \frac{2}{a} (1- \cosh (a \theta )) \Big)^2 - 4 (x^2 - t^2)} - 2 a^{-1} \sinh (a \theta ) \right)^3} , \nonumber \\
      E_{2} & = \frac{8 a y \left( x+ a^{-1} \right)}{ \left( \sqrt{\Big( 2 x + a ( -t^2 + x^2 + y^2 + z^2 ) + \frac{2}{a} (1- \cosh (a \theta )) \Big)^2 - 4 (x^2 - t^2)} - 2 a^{-1} \sinh (a \theta ) \right)^3} , \nonumber \\
      E_{3} & = \frac{8 a z (x+ a^{-1})}{ \left( \sqrt{\Big( 2 x + a ( -t^2 + x^2 + y^2 + z^2 ) + \frac{2}{a} (1- \cosh (a \theta )) \Big)^2 - 4 (x^2 - t^2)} - 2 a^{-1} \sinh (a \theta ) \right)^3} ,
\end{align}
and 
\begin{align} \label{small_angle_b2}
      B_{1} & = 0, \nonumber \\
      B_{2} & = -\frac{8 a z t}{\left( \sqrt{\Big( 2 x + a ( -t^2 + x^2 + y^2 + z^2 ) + \frac{2}{a} (1- \cosh (a \theta )) \Big)^2 - 4 (x^2 - t^2)} - 2 a^{-1} \sinh (a \theta ) \right)^3}, \nonumber \\
      B_{3} & = \frac{8 a y t}{ \left( \sqrt{\Big( 2 x + a ( -t^2 + x^2 + y^2 + z^2 ) + \frac{2}{a} (1- \cosh (a \theta )) \Big)^2 - 4 (x^2 - t^2)} - 2 a^{-1} \sinh (a \theta ) \right)^3},
\end{align}
Furthermore in the limits $a\to 0$ and $\phi\to 0$ we have that the velocity of the charge is also vanishing
\begin{align}
    \lim_{\phi \rightarrow 0, a \rightarrow 0} v^{i} (\theta) = \lim_{\phi \rightarrow 0, a \rightarrow 0} \left( \frac{\sinh a \theta}{\cosh a \theta \cosh \phi}, \ \frac{\sinh \phi}{\cosh \phi}, \ 0 \right) = \left( 0, 0, 0 \right).
\end{align}
As the result the expression for the electric field reduces to its standard Coulomb form:
\begin{align} \label{small_angle_e2_a0}
    \lim_{\phi \rightarrow 0, a \rightarrow 0} \vec{E} = \frac{ \vec{R}}{\left( t - \theta \right)^3}, \quad t-\theta = |\vec{R}|.
\end{align}
Meanwhile the magnetic field is vanishing.
As it should be, if $\phi$ is set to zero

If we consider the limit of zero acceleration keeping $\phi \neq 0$, then both electric and magnetic fields will be non-zero
\begin{align}
\begin{split}
    & \vec{E} = \frac{ \left( -x \cosh (\phi ), t \sinh (\phi )-y \cosh (\phi ), -z \cosh (\phi ) \right)}{\Big( \theta -t \cosh (\phi )+y \sinh (\phi ) \Big)^3}, \\
    &  \vec{B} = \frac{ \left( -z \sinh (\phi ), 0, x \sinh (\phi ) \right)}{\Big( \theta -t \cosh (\phi )+y \sinh (\phi ) \Big)^3}.  
\end{split}
\end{align}
We think that these observations clarify the properties of the solutions of the Maxwell's equations that we consider in this paper.

\section{Conclusions and acknowledgments}

Thus, we observe that a uniformly accelerating charge does create a radiation in the wave zone (i.e. as seen by a distant observer in Minkowski frame). At the same time in static zone corresponding to the co-moving Rindler's frame the radiation is absent. 

The question that remains to be considered is the number 4 among the questions formulated in the Introduction section.

This work was supported by the Ministry of Science and Higher Education of the Russian Federation (agreement no. 075–15–2022–287).

\appendix

\section{Appendix} \label{app_A}
In this appendix we study the radiation flux in the Rindler's frame in some peculiar situations. We observe that sometimes the flux is not zero and even is negative. 

Namely, we consider the behavior of the Poynting vector and the energy flux in the case where $\rho$ remains finite and $y^2 + z^2 \sim e^{\tau}$ are taken to infinity. To observe somewhat anomalous behavior of the energy flux, let us consider separately two cases when in the last limit $y$ remains finite (recall that along this direction the charge is moving with constant velocity) and the the case when in the limit under consideration $z$ remains finite.

In the first case $z^2 \sim e^{\tau} \rightarrow \infty$.
Then the Poynting vector \eqref{PoyntingRind} in this limit takes the following form:
\begin{align}
      & S_{1}^{R} \approx -\frac{\sinh^2 (\phi ) \Big( \cosh (a \theta) - \sinh(a \theta) \cosh(\phi) \Big)^4}{\pi a^{2} \rho z^4 \Big(\sinh(a \theta) - \cosh(a \theta) \cosh (\phi ) \Big)^6}, \nonumber \\
      & S_{2}^{R} \approx \frac{ 2 \sinh^2 (\phi ) \Big( \cosh (a \theta) - \sinh(a \theta) \cosh(\phi) \Big)^4}{\pi a^{3} z^6 \Big(\sinh(a \theta) - \cosh(a \theta) \cosh (\phi ) \Big)^6} \, \times \nonumber \\
      & \times \Big[a y + \coth (\phi ) \left( \cosh (a \theta) - \sinh(a \theta) \cosh(\phi) \right) \Big], \nonumber \\      
      & S_{3}^{R} \approx \frac{2 \sinh^2 (\phi ) \Big( \cosh (a \theta) - \sinh(a \theta) \cosh(\phi) \Big)^4}{ \pi a^{2} z^5 \Big(\sinh(a \theta) - \cosh(a \theta) \cosh (\phi ) \Big)^6},
\end{align}
and the energy flux ($|\vec{R} | \sim a z^2 $):
\begin{align}
    & \frac{dI^{R}}{d \Omega} = \Big(\vec{S}^R, \vec{n} \Big) \left|\vec{R}\right|^2 \sim \frac{ \sinh^2\left(\phi\right)}{a \rho z^2} \, \bigg( - a z^2 + o(z) \bigg),
\end{align}
where we used \eqref{BigDistanceRelation} to obtain these expressions. It can be seen that the flux in this limit does not tend to zero at infinity and even becomes negative for finite $\phi$. Please note, however, that the resulting flux is zero in the limit $\phi = 0$. And what is even more important is that the flux through the infinity depends on the spatial coordinate $\rho$, which is never the case in standard situations. Perhaps that is the hint for the resolution of the puzzle: the notion of the large spatial distance limit in the Rindler's frame is not that straightforward. 

For completeness let us also consider the second case mentioned at the beginning of the appendix. We see the same behavior of the energy flux in this case $y^2 \sim e^{\tau} \rightarrow \infty$:
\begin{align}
      & S_{1}^{R} = -\frac{\sinh^2 (\phi ) \Big( \cosh (a \theta) - \sinh(a \theta) \cosh(\phi) \Big)^4}{\pi a^2 \rho y^4 \Big(\sinh(a \theta) - \cosh(a \theta) \cosh (\phi ) \Big)^6}, \nonumber \\
      & S_{2}^{R} = \frac{ 2 \sinh^2 (\phi ) \Big( \cosh (a \theta) - \sinh(a \theta) \cosh(\phi) \Big)^4}{\pi a^2 y^5 \Big(\sinh(a \theta) - \cosh(a \theta) \cosh (\phi ) \Big)^6}, \nonumber \\      
      & S_{3}^{R} = \frac{2 z \sinh^2 (\phi ) \Big( \cosh (a \theta) - \sinh(a \theta) \cosh(\phi) \Big)^4}{ \pi a^2 y^6 \Big(\sinh(a \theta) - \cosh(a \theta) \cosh (\phi ) \Big)^6},
\end{align}
and the energy flux is ($|\vec{R}| \sim a y^2 $):
\begin{align}
    & \frac{dI^{R}}{d \Omega} = \Big(\vec{S}^R, \vec{n} \Big) \left|\vec{R}\right|^2 \sim \frac{ \sinh^2\left(\phi\right)}{a \rho y^2} \, \bigg( - a y^2 + o(y) \bigg).
\end{align}
Thus, it has the same properties as the flux of the first case.

\section{Appendix} \label{app_B}
To obtain equation \eqref{rindler_a_0}, consider the relation \eqref{txyz1a_rindler}:
\begin{align} \label{connection_rindler}
    & \rho (\cosh (\tau) \cosh (a \theta) - \sinh (\tau) \sinh (a \theta) ) = \nonumber \\
    & = \frac{a}{2} \left( a^{-2}+\rho ^2+y^2+z^2 \right) - \sinh (a\theta) (y \sinh (\phi) + \rho \sinh (\tau ) (1 - \cosh (\phi))).
\end{align}
Using this relation, we can express $\rho$ via other variables in the limit $a \rightarrow 0$:
\begin{align} \label{limit_a_0_rho}
    & \lim_{a \rightarrow 0} \rho = \lim_{a \rightarrow 0} e^{\pm \tau} \left( \frac{1}{a} \mp \theta \cosh(\phi) \right).
\end{align}
Since $\theta$ is the proper time, then from the above limit we obtain that at a finite time $\tau$ the spatial position of the radiation observation point must be at infinity. 

Reversing the argument, the eq. \eqref{connection_rindler} can be solved for $\tau$. Then for a finite $\rho$ the time of the radiation observation $\tau$ must tend to $\pm \infty$:
\begin{align} \label{limit_a_0_tau}
    & \lim_{a \rightarrow 0} e^{\tau} = \lim_{a \rightarrow 0} \rho^{-1} \left( \frac{1}{a} + \theta \cosh(\phi) \right) \text{ or } \lim_{a \rightarrow 0} e^{\tau} = 0.
\end{align}
Thus, in the case \eqref{limit_a_0_rho}, in the limit $a \rightarrow 0$ the four-potential in the Rindler region \eqref{a_in_rindler} will tend to
\begin{align}
      & \lim_{a \rightarrow 0} A_{0}^{R} (\tau, y, z) = \lim_{a \rightarrow 0} \frac{\rho \cosh (\tau ) \cosh (\phi ) }{\rho \sinh (\tau) \cosh (\phi ) - y \sinh (\phi )} = \coth(\tau), \nonumber \\
      & \lim_{a \rightarrow 0} A_{1}^{R} (\tau, y, z) = \lim_{a \rightarrow 0} \frac{y}{\rho} \cdot \frac{ \sinh (\phi )}{\rho \sinh (\tau) \cosh (\phi ) - y \sinh (\phi )} = 0, \nonumber \\
      & \lim_{a \rightarrow 0} A_{2}^{R} (\tau, y, z) = \lim_{a \rightarrow 0} -\frac{\sinh (\phi ) }{\rho \sinh (\tau) \cosh (\phi ) - y \sinh (\phi )} = 0, \nonumber \\
      & \lim_{a \rightarrow 0} A_{3}^{R} (\tau, y, z) = 0,
\end{align}
which agrees with (\ref{rindler_a_0}).
Such a four-potential is a pure gauge with $\alpha = \ln (\sinh(\tau))$. 

If we consider the other two cases \eqref{limit_a_0_tau}, the four-potential will tend accordingly
\begin{align}
    \lim_{a \rightarrow 0} A^{R}_{\mu} (\tau \rightarrow -\ln (a \rho), \rho, y, z) = \left(1, 0, 0, 0 \right),
\end{align}
or
\begin{align}
    \lim_{a \rightarrow 0} A^{R}_{\mu} (\tau \rightarrow -\infty, \rho, y, z) = \left(-1, 0, 0, 0 \right).
\end{align}
The resulting four-potentials are also pure gauges.

Let us consider also the behavior of a four-vector potential in the limit $a \rightarrow 0$ in the Minkowski coordinates. From the relation \eqref{txyz1a} we obtain the following limiting relation for the finite $t$:
\begin{align} \label{limit_a_0_x}
    & \lim_{a \rightarrow 0} x = \lim_{a \rightarrow 0} \left[ \frac{1}{a} \cosh(a \theta) \pm  \sqrt{ \theta^2 - 2 \theta (t \cosh(\phi) - y \sinh(\phi)) - (y^2 + z^2 - t^2) }\right],
\end{align}
and the another relation for the finite $x$:
\begin{align} \label{limit_a_0_t}
    & \lim_{a \rightarrow 0} t = \lim_{a \rightarrow 0} \left[\theta \cosh(\phi) \pm \left( \frac{1}{a} - x \right)\right].
\end{align}
Then the limiting values of the four-vector potential, when $a\to 0$, in the Minkowski frame in the cases \eqref{limit_a_0_x} and \eqref{limit_a_0_t} are, correspondingly:
\begin{align}
    & \lim_{a \rightarrow 0} A_{\mu} (x \rightarrow a^{-1} \cosh(a \theta), t, y, z) = \left( \cosh(\phi), 0,\frac{\sinh(\phi)}{ \theta - t \cosh(\phi) + y \sinh(\phi)}, 0 \right), \nonumber \\
    & \lim_{a \rightarrow 0} A_{\mu} (t \rightarrow \pm a^{-1}, x, y, z) = \left( 0, 0, 0, 0 \right).
\end{align}
It can be seen that in the first limit the four-vector is a pure gauge only when $\phi = 0$.




\end{document}